\documentclass[11pt]{combine}
\usepackage{amsmath,amsfonts,amssymb,amsthm,epsfig, graphicx, hyperref}
\usepackage[dvipsnames,table,dvipsnames*, svgnames*, hyperref]{xcolor}
\usepackage{natbib,amsmath,amssymb,amsthm,graphicx,setspace,paralist,booktabs,rotating,subcaption,float,color}
\usepackage{multirow}
\usepackage{fancyhdr}
\usepackage{bibunits}
\usepackage[small]{titlesec}

\newtheorem{theorem}{Theorem}
\newtheorem{corollary}{Corollary}
\newtheorem{proposition}{Proposition}

\newtheorem{lemma}{Lemma}
{
\theoremstyle{definition}
\newtheorem{definition}{Definition}
\newtheorem{example}{Example}

}
\newcommand{\beq}{\begin{equation}}
\newcommand{\eeq}{\end{equation}}
\newcommand{\beas}{\begin{align*}}
\newcommand{\eeas}{\end{align*}}
\newcommand{\bea}{\begin{align}}
\newcommand{\eea}{\end{align}}
\newcommand{\bei}{\begin{itemize}}
\newcommand{\eei}{\end{itemize}}
\newcommand{\ben}{\begin{enumerate}}
\newcommand{\een}{\end{enumerate}}
\newcommand{\bet}{\begin{theorem}}
\newcommand{\eet}{\end{theorem}}
\newcommand{\bel}{\begin{lemma}}
\newcommand{\eel}{\end{lemma}}
\newcommand{\bep}{\begin{proposition}}
\newcommand{\eep}{\end{proposition}}

\newcommand{\bed}{\begin{definition}}
\newcommand{\eed}{\end{definition}}
\newcommand{\bec}{\begin{corollary}}
\newcommand{\eec}{\end{corollary}}
\newcommand{\bex}{\begin{example}}
\newcommand{\eex}{\end{example}}

\numberwithin{equation}{section}

\newcommand{\T}[1]{{\text{T}}}

\usepackage{algorithm,algorithmic}

\begin{document}


\title{\scshape Missing at Random or Not: A Semiparametric Testing Approach}
\author{Rui Duan$^1$, C. Jason Liang$^2$, Pamela Shaw$^1$, Cheng Yong Tang$^3$ and Yong Chen$^1$ \\
$^1$Department of Biostatistics, Epidemiology and Informatics, \\
University of Pennsylvania,
Philadelphia, PA 19104\\
$^2$National Institute of Allergy and Infectious Diseases, Rockville, MD, 20852\\
$^3$Department of Statistical Science, Temple University, Philadelphia, PA 19122\\
}
\date{}
\maketitle
\thispagestyle{empty}

	\begin{abstract}
Practical problems with missing data are common, and statistical methods have been developed concerning the validity and/or efficiency of statistical procedures. On a central focus, there have been longstanding interests on the mechanism governing  data missingness, and   correctly deciding  the appropriate mechanism is crucially relevant for conducting proper practical investigations.  The conventional notions include the three common potential  classes -- missing completely at random, missing at random, and missing not at random.  In this paper, we present a new hypothesis testing approach for deciding between missing at random and missing not at random.  Since the potential alternatives of missing at random are broad, we focus our investigation on a  general class of models with instrumental variables for data  missing not at random.  Our setting is broadly applicable, thanks to that the model concerning the missing data is nonparametric, requiring no explicit model specification for the data missingness. The foundational idea is to develop appropriate discrepancy measures between estimators whose properties significantly differ  only when missing at random does not hold.  
We show that our new hypothesis testing approach achieves an  objective data oriented  choice between missing at random or not.  We demonstrate the feasibility, validity, and efficacy of the new test by theoretical analysis,  simulation studies,  and a real data analysis.
	\bigskip
	
\end{abstract}
\section{Introduction}


Missing data are common in almost all data collection procedures where some units and/or items are not observed  due to various reasons. How to appropriately deal with missing data has been a pervasive  and crucial consideration for drawing valid  scientific conclusions with statistical efficiency; see, among others, the monograph by \cite{Little2019}. 

In healthcare research, missing data are almost unavoidable, while their potential to undermine the validity of research results cannot be ignored. For example, the electronic health records (EHRs) data, as a result of not having been collected specifically for research purposes,  are subject to considerable missing data. Missing data can be due to a lack of collection (e.g., patient was never asked about a condition) or a lack of documentation (e.g., patient was asked about a condition but the response was never recorded in the medical record). Lack of documentation is particularly common when it comes to a patient not having a symptom/comorbidity. Instead of recording a negative value for each potential symptom/comorbidity, all data fields are left blank (missing) and only the positive values are recorded. Thus it can be difficult to differentiate between the lack of a comorbidity, the lack of documentation of a comorbidity and the lack of data collection regarding the comorbidity.  On the other hand, failure to account for the missing data can have a significant effect on the research conclusions \citep{sterne2009multiple,little2012prevention}.

A central methodological concern in the missing data literature is on the mechanism governing the data missingness. 
There have been intensively developing investigations in this area,  methodologically, theoretically, and practically.  Incorporating different types of the  data missingness mechanism is crucially relevant for developing appropriate statistical methods. 
In particular, the notions of missing completely at random, missing at random, and missing not at random are the commonly used classes for mechanisms characterizing the data missingness; see \cite{rubin1976} and 
\cite{Little2019}.  Missing completely at random refers to the case that  events leading to any particular data being missing are independent of both observed and unobserved variables. Such a case is simple, but rarely happens in practice. 

Missing at random supports a popular class of devices for investigating missing data problems.  It conceptually refers to that the data missingness depends on observed variables alone but not the unobserved ones. A great deal of effort has been devoted in methodological development under this case; and many approaches have become routines for solving practical problems. Foremost,  missing at random is convenient.  For some cases in this scenario,  applying  existing methods with no adjustment by  simply ignoring the missing data may lead to valid results; concerns are then on the efficiency and valid variance estimations for statistical inference; see, for example, \cite{White2010} and \cite{Bartlett2014}.  
Missing at random is meritorious for providing  unified  statistical frameworks.  That is,  approaches of standard form can be developed and applied in broad class studies. 
For example, the likelihood approaches, Bayesian approaches, imputation methods are abundantly available;  see the monographs by \cite{Molenberghs2007} and \cite{MFKTV_2014} for overview. 
For semiparametric methods using the generalized estimating functions \citep{LiangZeger_1986_Bioka},  as another class of examples, if data are missing at random and the missing propensity function is appropriately modeled and estimated, the inverse probability weighting approach can ensure the validity of the resulting inference.  For achieving the semiparametric estimation efficiency bound with missing data, one may design approaches by using the augmented inverse probability weighting approaches; see \cite{RRZ_1995_JASA},   \cite{Tsiatis_2006}, \cite{KimShao_2013}  and references therein.  


Missing not at random is most general; it refers to the case where the missingness  is allowed to depend on the variables that are missing. Problems with  missing not at random are more challenging.   A main reason  one may imagine is that there are plenty of individual possibilities associated with data missing not at random in various scenarios.  Practically, these problems require dedicated effort and they are typically solved individually with relatively smaller class of situations  compared with those under missing at random.  As a consensus, when data are missing not at random, the propensity model plays a more important role, so that it is a fundamental part of the model; see the overview in monographs by \cite{Molenberghs2007} and \cite{MFKTV_2014}.  

For addressing challenging tasks, handling data missing not at random requires extra support  from either stronger model assumptions and/or data structural information. 
For example, \cite{KottChang_2010} proposed some calibration approach in the context of survey sampling utilizing known totals of some variables; \cite{KimYu_2011_JASA} investigated semiparametric regression with an exponential  tilting modeling containing unknown parameters whose estimation requires extra information. 
It is even more challenging that these assumptions often cannot be validated without extra information and/or additional data collection procedure such as following-up studies. 

Recently, a class of methods by using the instrumental variables are actively developing.  
Here we note that the instrumental variables are respecting to the data missing mechanism, whose objective differs in some ways from the conventional  notion of instrumental variables as in econometrics and other areas where the data model is of the major concern. 
Among them,    \cite{Wang2014} considered generalized methods of moments, and handled missing not at random with some estimating functions utilizing instrumental variables. 
\cite{ZhaoShao_2015_JASA} proposed a novel approach for estimating generalized linear models under data missing not at random, without requiring the specification of a model for missing data mechanism.  \cite{ShaoWang_2016} incorporated instrumental variables in inverse probability weighting, and relaxed requirement of extra information for the semiparametric approach of 
\cite{KimYu_2011_JASA}.   
\cite{Riddles2016} developed a propensity score adjusted likelihood approach.    \cite{Miao2016} investigated double robustness of the estimations with instrumental variable approaches. 
\cite{Morikawa2017} studied optimality of the estimation in a setting with instrumental variables.  \cite{Zhao2018} constructed optimal pseudo-likelihood approach with data missing not at random. 
\cite{Wang2019} recently considered  propensity selection and data modeling with some penalized information criteria. \cite{Tchetgen2018} studied identifiability in semiparametric estimation with instrumental variables; see also \cite{Miao2018}.





Usually with missing not at random, correct specification of a model for the data missingness mechanism is required to ensure valid and/or efficient inference. 
However, such a specification is generally difficult as it closely depends on the part of the data that are not observed. In addition, methods to address missing not at random can appreciably increase the complexity and the uncertainty of study results;  so that its validity is clearly more desirable. 
Nevertheless, 
relative to the recent surge of the development in studying missing not at random,  few methodology is available  for conducting statistical testing against the propensity model specifications. 
\cite{Mohan2014} considered testing for data missing mechanism in the context of directed acyclic graph and causal relationships between variables; they established conditions when such testing is possible.   Recently, \cite{Breunig2019} proposed a method with instrumental variables for testing the missing at random assumption with some squared integrated distance built upon some conditional moment identities.


We consider in this study a new approach for objectively deciding the mechanism of the data missingness from the two classes: missing at random or missing not at random. Such an approach has practical interest to an applied scientist who can be reluctant, without concrete supporting evidence, to undertake analysis methods for missing not at random. 
Inspired by recent development in handling missing not at random with instrumental variables, we  develop a semiparametric hypothesis testing  framework with this class of methods. 
To stay focused in our presentation, we consider the generalized linear models  \citep{GLM_1989_MN}, while recognizing the principle of the development broadly applies.  
For the mechanism governing the data missingness, we are motivated by the setting of \cite{ZhaoShao_2015_JASA} where the specification of a model for the missing data mechanism is not required, entitling broad validity of the resulting estimator. 
%
Then 	inspired by the rationale of the famous Hausman's test  \citep{hausman1978specification}, we propose  to identify two parameter estimators that both are valid if missing at random holds, and only one is valid otherwise.    	Then the signal for our test relies on a  discrepancy measure
taking significantly larger values when missing not at random is more appropriate. 
There are many candidates for the first estimator, and in our study we take the inverse probability weighted estimator as a concrete example in our development; and we observe that our method broadly applies.  
For the second estimator, we apply  the semiparametric approach of \cite{ZhaoShao_2015_JASA} that is valid for both classes of data missingness. 
%
Our theory confirms that the testing procedure is valid and powerful, and our simulation studies show that the test works satisfactorily when assuming data missing at random is not appropriate.

Our investigation makes a few contributions. 
Practically, we provides a framework that missing at random assumption can be tested against a broad setting.  Then to what end such an assumption is reasonable can actually be evaluated.  
Methodologically, our approach attempts to adequately exploits the current setting for data missing not at random, and we  demonstrate that testing the missing at random assumption is possible. 
As for technical development, existing Hausman's tests are developed with full parametric models, while our development extends their applicability to broad semiparametric settings. We analytically calculate the influence function of the semiparametric estimator of \cite{ZhaoShao_2015_JASA}, which provide a tool for broad statistical inferences. 
Our development accommodating  a nonparametric component  as an infinite dimensional nuisance parameter is a new feature of its own interest in the context of  model specification tests in the context of missing data problems. 

%
%
The rest of this article is organized as follows. In Section 2, we propose a general testing approach and provide a specific test statistic as an example. We showed the limiting distribution of the test statistic is a $\chi^2$ distribution.  In Section 3, we apply the proposed test to data from the Keep-it-off study to investigate the missing mechanism of the self-reported body weight of participants in a weight loss program.  In Section 4, we present a comprehensive simulation study to show the performance of the proposed test in terms of type I error and power.  We provide a discussion in Section 5, and we include the regularity conditions and part of the proofs in the Appendix.

\section{Main Development}\label{s2}

\subsection{Setting and Assumptions} 

We concretely consider a generic setting of regression analysis in this study.  
For a data set with a response variable $Y$ and covariates $X = (U^{T}, Z^{T})^{T}$ (the difference between $U$ and $Z$ to be explained later), we consider a setting with generalized linear model. Specifically, the conditional distribution of $Y$ given covariates $X$ belongs to the exponential dispersion family, and the conditional mean of $Y$, $\mu=E (Y|X)$, is related to covariates through a link function $h(\cdot)$, i.e.
\begin{equation}\label{model}
p(y|x; \theta) = \exp \left\{\frac{y\eta-b(\eta)}{\lambda}+c(y;\lambda)\right\}, \ \ \ h\{\mu(\eta)\} = \alpha +u^{T}\beta_u +z^{T}\beta_z,
\end{equation}
where $\eta$ is the natural parameter, $\lambda$ is the dispersion parameter, $\beta = (\alpha, \beta_u^{T}, \beta_z^{T})^{T}$ are regression coefficients, and $\mu(\eta)=b'(\eta)$ by the property of the exponential family. We denote the dimensions of $U$, $Z$ and $X$ by $m_u$, $m_z$ and $m$ respectively.

As in econometrics literature \citep{Heckman1979,Heckman1997}, instrumental variables originally refer to those satisfying two conditions: 1)   they are conditional independent of the outcome variables that can be missing, given other variables; and 2) they are affecting the missing mechanism in a model with all variables.   In recent missing data literature,  
conditional independence is assumed between the instrumental variables and the missingness given all other variables including the missing ones.  Additionally,  some identifiability  conditions are needed for the parametric data model   on the connections between the instrumental variables and the response variables; see, for example, \cite{Fang2016} and \cite{Wang2019}. 

We consider that the response variable $Y$ can be missing, and let $R_i=1$ if $Y_i$ is observed, and $0$ otherwise.
We consider the following two assumptions as the possible underlying data  missingness mechanisms. \\
{\textrm{Assumption A1:}} \quad $P(R|Y, U, Z)=P(R|U)$;\\
{\textrm{Assumption A2:}} \quad $P(R|Y, U, Z)=P(R|Y, U)$.


Assumption A1 refers to the case of missing at random, where the missingness  is conditional independent of $Y$ that can be missing. 
Assumption A2 corresponds to the case of missing not at random. Here the component $Z$ is referred to as the instrumental variables. 
Clearly,  A1 is stronger, and itis a special case of A2. Thus, we note that a valid estimator under A2 remains valid under A1. Here common to both assumptions is that,  conditioning on $Y$ and $U$, the variable $Z$ is independent of the missing data indicator $R$. 

Estimating model parameters with missing at random has been intensively studied, and  commonly applied approaches include the inverse probability weighting, augmented inverse probability weighting, multiple imputations and others;  see \cite{Tsiatis_2006}, \cite{ Molenberghs2007}, and \cite{Little2019}. 
When data are missing not at random, inference on $\beta$ commonly requires specification of the missing propensity function. Nevertheless, with the instrumental variable $Z$, valid estimation of the parameter $\beta$ can be made semiparametrically without a need to specify a model for the missingness propensity function \citep{ZhaoShao_2015_JASA}, subject to some identifiability conditions. 

We now define some notations. 
For any function $t(\theta,X,Y,R)$, we use $\nabla_\theta t(\theta,X,Y,R) = \partial t(\theta,X,Y,R)/\partial \theta$, and $\nabla_{\theta\theta} t(\theta,X,Y,R) = \partial^2 t(\theta,X,Y,R)/\partial \theta^2$. We understand $t_i$ as $t(\theta,X,Y,R)$ with $(X,Y,R)$ replaced by $(x_i,y_i,r_i)$, and treat $t_i$ as a random variable $(i=1,\dots,n)$.

\subsection{Methodology}
In practice, it is important to understand what the true underlying missing data mechanism is, which can be formulated as testing Assumption A1 against Assumption A2.  
Since A1 is a special case of A2,  a rationale to develop a statistical test is then identifying two estimators for $\beta$, $\widehat \beta$ and $\widetilde \beta$,   such that $\widehat\beta$ is valid  only under assumption A1, and $\widetilde\beta$ is valid under assumption A2. 
Clearly,   $\widehat\beta$ is expected to be biased when A1 is violated; and $\widetilde\beta$ is  valid under both A1 and A2.  
Subsequently, the discrepancy between  $\widehat \beta$ and $\widetilde \beta$ becomes the signal for detecting violation of Assumption A1. When such discrepancy is large, we have evidence that missing at random is suspicious. Such a rationale is the spirit of the classical Hausman’s test \citep{hausman1978specification}, though the test was developed for parametric estimators and did not formally address estimators that require nonparametric estimation of nuisance parameters.

In our framework, a main challenge is that we have a broad assumption of the form A2.  Here we intend to impose minimal restriction on the form of the propensity function -- allowing it to be nonparametric -- so that our approach is broadly applicable.  
It is worth to note that that conventional Hausman's test does not apply due to this scope of our work incorporating  a setting requires handling  a nonparametric distribution.   
Indeed, as shown in our later development,  profiling out this nonparametric component as a functional nuisance parameter involves major technical challenges for constructing the testing statistic.  
Therefore, as an interest of its own, our test statistic extends the scope of the classical   Hausman's test with nonparametric nuisance parameters that can be infinite dimensional. 


Concretely, we first show the following theorem  describing the asymptotic property of the discrepancy measure $(\widehat\beta-\widetilde\beta)$ in a general setting accommodating semiparametric estimations. 
We denote by $\Lambda_1$ and $\Lambda_2$ nuisance parameters  respectively under Assumption A1 and A2 of the specific  approaches for estimating the model parameter $\beta$.    Since model estimation approaches are broad including those parametric and semiparametric ones, here $\Lambda_1$ and $\Lambda_2$ are allowed to be general including functional-valued quantities such as  nonparametric propensity functions and distribution functions in some semiparametric models. They essentially depend on the model settings under which the estimation methods are developed; see our examples in Section \ref{E1} and \ref{E2}.

\begin{theorem} \label{tm:0}
	Under Assumption A1, suppose the estimator $\widehat\beta$ satisfies
	\[
	\sqrt{n}(\widehat\beta-\beta_0) = \frac{1}{\sqrt{n}}\sum_{i=1}^{n}\psi(y_i,x_i;
	\beta_0,\Lambda_1) +o_p(1),
	\]
	for some function $\psi(y_i,x_i;
	\beta_0,\Lambda_1)$, 
	and under Assumption A2, the estimator $\widetilde\beta$ satisfies
	\[
	\sqrt{n}(\widetilde\beta-\beta_0) = \frac{1}{\sqrt{n}}\sum_{i=1}^{n}\phi(y_i,x_i;
	\beta_0,\Lambda_2) +o_p(1),
	\]
	for some function $\phi(y_i,x_i;
	\beta_0,\Lambda_2)$.  Assume $E \{\psi(y_i,x_i;
	\beta_0,\Lambda_1)\} = 0$, $E \{\phi(y_i,x_i;
	\beta_0,\Lambda_2)\} = 0$,  $E\{ \psi(y_i,x_i;
	\beta_0,\Lambda_1)\psi(y_i,x_i;
	\beta_0,\Lambda_1)^T\} \le \infty$, and $E\{ \phi(y_i,x_i;
	\beta_0,\Lambda_2)\phi(y_i,x_i;
	\beta_0,\Lambda_2)^T\}\le \infty$. Then we have that under Assumption A1, 
	the discrepancy measure
	$
	(\widetilde{\beta}-\widehat{\beta}) 
	$ satisfies
	\[
	\sqrt{n}(\widetilde{\beta}-\widehat{\beta})\rightarrow N(0, W),
	\]
	where $W = E \left[\{\psi(y_i,x_i;
	\beta_0, \Lambda_1)-\phi(y_i,x_i;
	\beta_0, \Lambda_2)\}\{\psi(y_i,x_i;
	\beta_0, \Lambda_1)-\phi(y_i,x_i;
	\beta_0, \Lambda_2)\}^T\right]$. 
\end{theorem} 

In Theorem \ref{tm:0}, $\psi(\cdot)$ and $\phi(\cdot)$ are known as the influence functions.  
Theorem \ref{tm:0} ensures that  to construct a test statistic, one needs to find  a consistent estimator for the variance matrix $W$. 

In existing Hausman's tests,  if $\widehat{\beta}$ is efficient using the likelihood approach under Assumption A1, the variance matrix $W$ has the property that $W = V_2-V_1$, where $V_2$ is the variance of $\widetilde{\beta}$ and $V_1$ is the variance of $\widehat{\beta}$. In such a case,  the application of the  Hausman's test is straightforward. 

More generally,  however, there are two major difficulties in our study.   First, choices of the estimators are broad, and semiparametric approach are popular in existing methods.  Thus, calculating the variances of $\widehat \beta$ and $\widetilde \beta$ is generally involved due to nonparametric nuisance parameters. Second, our framework accommodates cases when $\widehat \beta$ is not necessarily efficient. Thus $\widehat \beta$ and $\widetilde \beta-\widehat \beta$ could be correlated, and it no longer holds with a simple form that  $W = V_2-V_1$.   To tackle this challenge, we propose to apply the sample covariance matrix of the difference between the influence functions, i.e.,   $\psi(y_i,x_i;
\beta_0, \Lambda_1)-\phi(y_i,x_i;
\beta_0, \Lambda_2)$, $(i=1,\dots, n)$  as the estimator for $W$.  With a consistent estimator $\widehat W$, we can show that the test statistic $T = n(\widehat\beta-\widetilde\beta){\widehat W}^{-1}(\widehat\beta-\widetilde\beta)$ converges in distribution to $\chi^2_{m+1}$.


\subsection{Estimator Under Missing at Random -- Inverse Probability Weighting}\label{E1}

There are broad choices for estimating the model parameter when data are missing at random. 
In this case, even the complete-case analysis ignoring all missing data may provide a valid inference \citep{White2010}. To make our method more generalizable, we consider here an exemplary method for missing at random data, which is the inverse probability weighting (IPW) method with nonparametric estimation of the  propensity function.  

When the true propensity $\pi_0(u) = P(R = 1|U = u)$ is known, the inverse probability weighted likelihood  score equation can be constructed  as
\begin{equation}\label{EEMAR1}
g_n(\beta)=\frac{1}{n} \sum_{i=1}^{n} \frac{r_i}{\pi_0(u_i)}S(y_i,x_i;\beta),
\end{equation}
where $S(Y,X;\beta)= {\nabla_{\beta}\log p(Y|X;\beta)}$. However, in practice, the  propensity function is  unknown. In this case, parametric models can be specified for the missing propensity, and a two-stage method can be used to obtain a plug-in estimating equation. To avoid misspecification of the propensity function, we consider estimating $\pi_i$ through nonparametric regression. Let $J(x)$ be a multivariate kernel function of variable $u$ satisfying $\int J(u) du = 1$. The missing propensity $\pi_0(u)$ can be estimated by a kernel regression estimator of form
\begin{equation*}
\widehat {\pi}_b(u) = \frac{\sum_{i=1}^{n}r_iJ(\frac{u-u_i}{b})}{\sum_{i=1}^{n}J(\frac{u-u_i}{b})},
\end{equation*} 
where $b$ is the bandwidth that determines smoothness of $\widehat{\pi}_b$. By plugging in the estimator  $\widehat {\pi}_b(u)$ back to estimating equation~(\ref{EEMAR1}), we obtain the following estimating equation
\begin{equation}\label{EEMAR2}
\widehat{g}_n(\beta) = \frac{1}{n} \sum_{i=1}^{n} \frac{r_i}{\widehat {\pi}_b(u_i)}S(y_i,x_i;\beta) = 0.
\end{equation}

With a proper choice of bandwidth $b$, it can be shown that the kernel regression estimator $\widehat {\pi}_b(u)$ is a consistent estimator of $\pi_0(u)$. Although the convergence rate of $\widehat {\pi}_b(u)$ is slower than ${n}^{1/2}$ \citep{hardle1993comparing}, as suggested by \cite{Newey_1994} and \cite{NeweyMcFadden1994} the estimator obtained from the plug-in estimating equation is still  ${n}^{1/2}$-consistent and asymptotically normal. Specifically, we have the following theorem. 

\begin{theorem}\label{tm:1}
	Let $\widehat{\beta}$ be the estimator solving estimating equation~(\ref{EEMAR2}). Under regularity conditions R1--R4 in Appendix A, we have
	\begin{equation*}
	n^{1/2} (\widehat{\beta}-\beta_0) \rightarrow N(0,\Sigma), \quad with \quad \Sigma = E(\psi_i\psi_i^{T}),
	\end{equation*}
	as $n\rightarrow \infty$, where $\psi_i$ is an influence function of $\widehat{\beta}$, calculated as
	\begin{equation*}
	\psi_i = \frac{r_i}{\pi_0(u_i)}E\left[-\nabla_{\beta}{S(Y,X;\beta)} \right]^{-1}S(y_i,x_i;\beta).
	\end{equation*}
\end{theorem}
The proof of Theorem \ref{tm:1} is outlined in the Supplementary Material. 
This is an example of parameter estimation with nonparametric propensity-weighted score functions.
Here, it is the key to develop the influence function $\psi_i$  with the propensity function $\pi_0(u)$ estimated nonparametrically.  Our  methodological framework here also more broadly applies. 

\subsection{An Estimator Under Missing Not at Random }\label{E2} 

The missing not at random case is a more challenging scenario from multiple aspects; see, among others,  \cite{KimShao_2013}.
Our attempt here has a consideration to be most accommodative  with no extra assumption beyond A2.
We have such a candidate, as elaborated below.   
As pointed out by \cite{ZhaoShao_2015_JASA},  
a pseudolikelihood approach can be built with the instrumental variables $Z$ that satisfy Assumption A2. 
A key observation is the following equivalence between the conditional distributions
\begin{equation}\label{eq:00}
P(Z|Y,U,R=1) = P(Z|Y,U) = \frac{p(Y|U,Z;\beta)P(U,Z)}{\int p(Y|U,z;\beta)P(U,z)dz}, 
\end{equation}
where the first equation is from A2, and the second equation is simply its definition.  Hence, from (\ref{eq:00}) evaluating the likelihood is valid with complete only data such that $R=1$. 

To avoid a fully parametric specification of the model, the joint distribution $P(U,Z)$ becomes the nonparametric nuisance parameter -- a major difference from the development in Section \ref{E1} with missing at random. 
In this context, we 
consider the nonparametric product kernel estimator 
\begin{equation*}
\widehat f(x) = \frac{1}{n}\frac{1}{h^m}\sum_{i=1}^{n}\prod_{d=1}^{m}K\left(\frac{x_d-X_{di}}{h}\right) = h^{-m}\sum_{i=1}^{n}K_m\left(\frac{X-X_i}{h}\right), \quad x=(x_1,\dots,x_m)^T,
\end{equation*}
where $K$ is a one-dimensional kernel function,  $h$ is the bandwidth and $K_m$ denotes the $m$-dimensional kernel function. A plug-in semiparametric log-pseudolikelihood function is then obtained  
\begin{equation}\label{EEMAR3}
\ell(\beta,\widehat{F}) = \frac{1}{n}\sum_{i=1}^{n}R_i{\log\frac{p(y_i|u_i,z_i;\beta)\widehat f(u_i,z_i)}{\int p(y_i|u_i,z_i;\beta)\widehat f(u_i,z)dz} } = \frac{1}{n}\sum_{i=1}^{n}H_i(\beta,\widehat{F}),
\end{equation}
where $\widehat{F}$ is the corresponding empirical cumulative distribution function of $\widehat f$.  Then  the maximum pseudolikelihood estimator is $\widetilde{\beta}={\textrm{argmax}_{\beta}} \left\{ \ell(\beta,\widehat{F}) \right\}$. 
We also have that the first order derivative function
\[
h(\beta,\widehat{F}) = \frac{1}{n} \sum_{i=1}^{n}{\nabla_\beta H_i(\beta,\widehat{F})} =\frac{1}{n} \sum_{i=1}^{n}r_i\left\{ \frac{\nabla_{\beta}p(y_i|u_i,z_i;\beta)}{p(y_i|u_i,z_i;\beta)} -\frac{\int \nabla_{\beta}p(y_i|u_i,z;\beta)\widehat{f}(u_i,z)dz}{\int p(y_i|u_i,z;\beta)\widehat{f}(u_i,z)dz}\right\},
\] 
and let $\mu(F) = E\{\nabla_\beta H(\beta_0,F)\}$ for a given distributional  function $F$ of $U$ and $Z$.

To construct a discrepancy measure between two estimators $\widehat \beta$ and  $\widetilde \beta$,   both of their influence functions are required.  The influence function of $\widetilde \beta$ is more complicated, and its explicit form is not known in the literature. 
To solve this difficulty, 
we establish the following lemma,  providing the explicit form of the approximation error between the nonparametric estimation and the truth, which is a key component in the influence function of $\widetilde \beta$. 


\begin{lemma}\label{la:1}
	Assume regularity condition R9 holds.  Let 
	\begin{equation*}
	\delta_i =\int \pi(y,u_i)\frac{\int \nabla_\beta p(y|u_i,z;\beta)f_0(u_i,z)dz}{\int p(y|u_i,z;\beta)f_0(u_i,z)dz}p(y|u_i,z_i;\beta)dy-\int \pi(y,u_i)\nabla_\beta p(y|u_i,z_i;\beta)dy
	\end{equation*}
	where $\pi(u,y) = P(R = 1|y,u)$. Then 
	\[
	n^{1/2}\{\mu(\widehat{F})-\mu(F_0)\} = {n^{-1/2}}\sum_{i=1}^{n}\delta_i +o_p(1)
	\]  
	with $E\{\delta_i\}=0$ and $E\{\delta_i\delta_i^{T}\} < +\infty$.
\end{lemma} 
The proof of Lemma \ref{la:1} is outlined in Appendix B where appropriately handling the nonparametric joint distribution $P(U,Z)$ is the key. 
As an intermediate result for constructing our test, we present the following theorem, refining the results of  \cite{ZhaoShao_2015_JASA} by providing the asymptotic distribution of $\widetilde{\beta}$, as well as an explicit form of its influence function. 
\begin{theorem}\label{tm:2}
	Under regularity conditions R5--R9 in Appendix A, we have
	\begin{equation*}
	n^{1/2} (\widetilde{\beta}-\beta_0) \rightarrow N(0,\Omega), \quad with \quad \Omega = E(\phi_i\phi_i^{T})
	\end{equation*}
	as $n\rightarrow \infty$, where $\phi_i$ is the influence function of $\widetilde{\beta}$, calculated as
	\begin{equation*}
	\phi_i = -E\{\nabla_{\beta\beta} H(\beta_0,F_0)\}^{-1}\{\nabla_{\beta}H_i(\beta_0,F_0)+\delta_i\}
	\end{equation*}
	and $\delta_i$ is given in Lemma  \ref{la:1}. 
\end{theorem}
A proof is given in  Appendix B. The form of $\phi_i$ in Theorem \ref{tm:2} has clear insights.  The contribution in $\phi_i$ without $\delta_i$ reflects the accuracy of the oracle estimator as if the true distribution $F$ is known.  The part from $\delta_i$ reflects the approximation error due to the nonparametric estimation.    Although the nonparametric distribution estimation has a slower convergence rate than $n^{1/2}$, the parameter estimation still achieves the $n^{1/2}$ rate,  echoing the phenomenon known in the literature on semiparametric estimations; see, for example \cite{Newey_1994} and \cite{NeweyMcFadden1994}.

\subsection{Testing Missing at Random or Not}
With the two estimators obtained in Sections 2.3 and 2.4, we construct by evaluating the discrepancy between $\widehat \beta$ and $\widetilde \beta$ such that the test statistic is expected to take a large value when Assumption A1 is violated. Concretely, let $T$ be a test statistic defined as 
\begin{align}\label{eq:testt}
T = n (\widetilde{\beta}-\widehat{\beta})^{T}\widehat W^{-1}(\widetilde{\beta}-\widehat{\beta}),
\end{align}
where the appropriate weighting matrix $W$ is constructed from the estimated influence functions $\phi_i$ and  $\psi_i$ as in Theorems \ref{tm:1} and \ref{tm:2}: 
\begin{equation*}
\widehat W = {1 \over n} \sum_{i=1}^{n}\left\{\widehat\phi_i-\widehat\psi_i\}\{\widehat\phi_i-\widehat\psi_i\right\}^{T}. 
\end{equation*}
The estimators of $\phi_i$ and $\psi_i$ are respectively defined as
\begin{equation*}
\widehat\phi_i = \left\{{1\over n}\sum_{i=1}^{n}-\nabla_{\beta\beta} H_i(\widetilde\beta,\widehat{F})\right\}^{-1}\left\{\nabla_{\beta}H_i(\widetilde\beta,\widehat{F})+\widehat\delta_i\right\}, 
\end{equation*}
and 
\begin{equation*}
\widehat\psi_i = \left\{{1\over n}\sum_{i=1}^{n}-\nabla_{\beta} S(\widehat\beta,y_i,x_i)\right\}^{-1}\frac{r_i}{\widehat{\pi}_h(u_i)}S(\widehat\beta,y_i,x_i),
\end{equation*}
with
\begin{equation*}
\widehat\delta_i = \widehat{\pi}_h(u_i)\left[\int \frac{\int \nabla_\beta p(y|u_i,z;\widetilde\beta)\widehat f(u_i,z)dz}{\int p(y|u_i,z;\widetilde\beta)\widehat f(u_i,z)dz}p(y|u_i,z_i;\widetilde\beta)dy - \int \nabla_\beta p(y|u_i,z_i;\widetilde\beta)dy\right]. 
\end{equation*}
The key property of the test statistic $T$ in equation~(\ref{eq:testt}) is described in the following corollary.  
\begin{corollary} \label{tm:3}
	Under the null hypothesis that Assumption A1 is true, and under regularity conditions R1--R9 in Appendix A,  as $n\rightarrow \infty$,  the test statistic $T$ converges weakly to $\chi^2_{m+1}$.
\end{corollary} 
A proof is provided in the Supplementary Material. Corollary \ref{tm:3} reveals a central $\chi^2$ limiting distribution of $T$ under the null hypothesis, making it convenient for practical implementation of the proposed test.

We then evaluate the asymptotic power of the test statistic $T$ in equation~(\ref{eq:testt}) under a broad class of local alternatives. Suppose the missing propensity relates to $Y$ through a parameter $\gamma$, which is with finite dimension $q\ge 1$.  We denote the propensity by $P (R=1|y, u) = \pi(y,u;\gamma)$. Suppose when $\gamma = 0$, the missingness is at random and $\pi(y,u;\gamma)$ reduces to $\pi_0(u)$. Assume that $\pi(Y,U;\gamma)$ is second order differentiable, and there exist $\epsilon>0$ such that $ \pi(y,u;\gamma) \in (\epsilon,1-\epsilon)$ for all $\gamma$. Consider a fixed $\gamma_0$ and a sequence of alternatives:
\begin{equation*}
H_\alpha: \quad \gamma = n^{-1/2}\gamma_0
\end{equation*}
With LeCam's third lemma \citep{Van_1998}, we can establish the following results.

\begin{theorem}\label{tm:4}
	Under the alternative $H_\alpha$ and the same regularity conditions as in Theorem \ref{tm:3}, the limiting distribution of the test statistic $T$ is $\chi^2_{m+1}\{\gamma_0^{T}\eta^{T}W^{-1}\eta\gamma_0\}$, where $\eta = {Cov}(\phi_i-\psi_i, \zeta_i)$, and 
	\begin{equation*}
	\zeta_i = r_i \frac{\nabla_{\gamma}\pi(u_i,y_i,0)}{\pi_0(u_i)}-(1-r_i) \frac{\nabla_{\gamma}\pi(u_i,y_i,0)}{1-\pi_0(u_i)}. 
	\end{equation*}
\end{theorem}

A proof of the theorem is given in the Supplementary Material. The parameter $\gamma$ represents the dependence of the missingness propensity function on $Y$.  Theorem \ref{tm:4} implies that the test statistic remains bounded under the alternative $H_\alpha:\gamma = n^{-1/2}\gamma_0$, and its power is determined by the non-centrality parameter $\gamma_0^{T}\eta^{T}W^{-1}\eta\gamma_0$.   Furthermore, when the signal of missing not at random is not vanishing to zero at a rate faster than $n^{1/2}$, the power of the test with $T$ will go to one.

\section{Real data Analysis}

We analyzed data from the Keep It Off randomized controlled trial which recruited 191 participants and collected their body weight over a 1-year period \citep{shaw2017design}. Participants were randomized to receive one of three interventions to help maintain their weight after a recent weight loss. The primary outcome was the participant's weight at the in-person weigh-in at month six. As part of the follow-up, the participants reported their daily body weights through use of a wireless scale that transmitted data to the Way to Health portal, an online clinical study platform \citep{WTH2009}.  For more detailed information about the design of the trial, we refer to \cite{shaw2017design}.
In this example, we investigated the nature of the missing data for the daily at-home weigh-in data for the combined cohort, with the help of a regression analysis between daily weight and baseline covariate variables collected in the trial. 

\subsection{A gold standard outcome without missing}

We chose the self-reporting at-home weights at the day before the month-six visit as the outcome ($Y$). The missing proportion of the at-home weights is around 47\%. 
Since the primary outcome, the participant's weight at the in-person weigh-in at month six, was measured for all participants,
we considered it 
as a gold standard ($Y^\star$) since it is reasonable to assume no large variation of weight within two days.  The availability of the gold standard $Y^*$ in this scenarios allows us to validate our methods. 
\subsection{Select an instrumental variable with the help of the gold standard body weight}

We observed variables including age, gender, race, education-level, baseline weight, baseline boday mass index (BMI), treatment arms, and self-reported scores of physical activity and eating
habits based on a questionnaire provided by the Way to Health portal. In the existence of the gold standard in person weight at month six, we first fitted a linear regression using all the aforementioned variables as covariates. Among all these variables, only baseline weight and baseline BMI are significantly associated with weight at month six, and this finding is also consistent with many existing findings \citep{ben2003predictors}. 
Hence, to reduce the complexity of the proposed test, we chose to use the linear regression between daily weight and the baseline weight and BMI. 

Among  baseline weight ($U$) and BMI ($Z$), baseline BMI was hypothesized to be an instrumental variable, which should be (1) associated with the daily weight, and (2) after conditioning on the daily weight and the baseline weight, the missingness of the daily weight at a specific day should not be related to the baseline body mass index. To investigate  hypothesis (1), we fitted a linear regression between the gold standard $Y^*$ and $X = (U,Z)$ and the results were shown in Table \ref{linear}. We observed that the body mass is significantly associated with the outcome $Y^*$ (p-value $<$ 0.01). 

\begin{table}	
	\caption{Model fitting results for linear regression. The outcome is the weight at month 6, covariates  baseline weight, and baseline body mass index.} \label{linear}
	\centering
	\begin{tabular}{cccc}
		\hline
		& effect size &SE & P-value\\
		\hline
		intercept &9.06&7.66&0.23\\
		$U$: baseline weight & 1.07&0.04& $<0.01$\\
		$Z$: body mass index  & -0.76&0.30& $0.01$\\
		\hline
	\end{tabular}
\end{table}

To validate assumption (2), we fitted the following logistic regression 
\begin{equation}\label{mmnar}
\textup{logit} \{P(R=1)\} = c_0 +c_1 U +c_2 Y^* +c_3 Z,
\end{equation}
where $R = 1$ if $Y$ is observed. Specifically, Table \ref{table3} below showed the model fitting results of the above logistic regression, where the coefficient of the potential instrumental variable, baseline body mass index ($Z$), is statistically non-significant (p=0.26). The observation is reasonable that the current weight and baseline weight will capture information regarding current weight loss, likely a dominant determinant in whether a participant decides to weigh-in on a given day.

\subsection{Apply and validate the proposed method}

Applying our method using the baseline body mass index as an instrumental variable led to a test statistic of 9.12, compared to a chi-squared distribution with $3$ degrees of freedom, corresponding to a p-value of 0.02. This test suggested that the data missingness is not at random.  This matched the observation from the data that people having smaller body weight are more likely to report their weight for a given body mass index and baseline weight. This is consistent with the conventional wisdom that participants in weight loss trials that are gaining weight in the trial are disappointed at the lack of success and more likely to skip reporting.
\begin{figure}
	\caption{Plots of the data missingness propensity functions: left panel plots the propensity functions estimated respectively with missing at random or missing not at random assumptions and a linear fitting (red solid line) with 95\% pointwise confidence interval (dashed line); right panel plots the log odds ratio of the estimated propensity functions versus the gold standard weight with a curve (green) curve fitted by LOESS method \citep{harrell2015regression}. } 
	{\centering{
			\includegraphics[scale = 0.5]{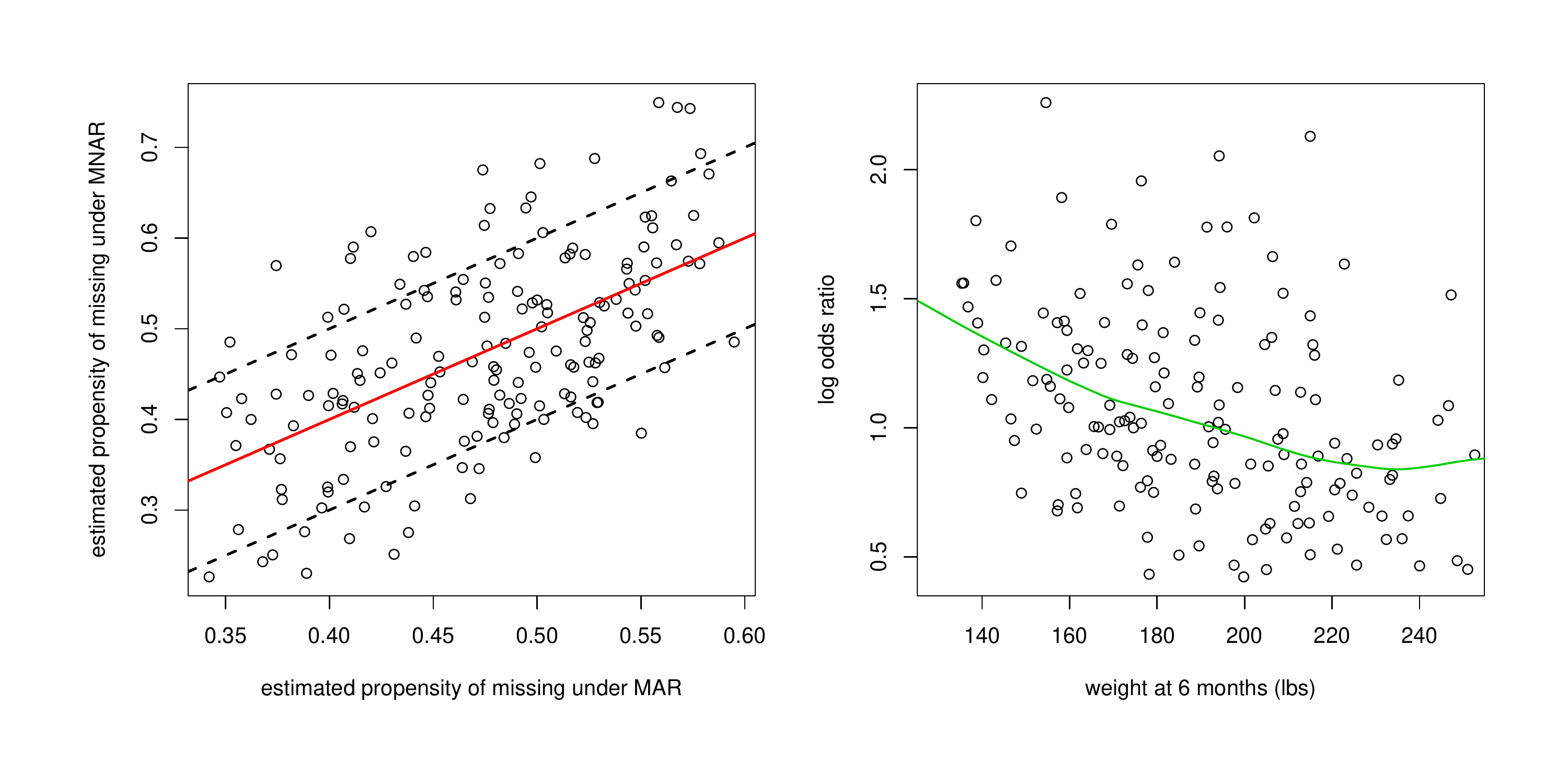}}}
\end{figure} 

This conclusion is also validated by the result in Table \ref{table3}. Specifically, we found that conditioning on baseline weight and body mass index, weight at month 6 ($Y^\star$), as an approximation to $Y$, remains statistically significantly associated with the missing indicator (p-value $=$ 0.02), which also suggested that the data are not  missing at random. 

Finally, Figure 1 provided a graphical evidence of the missing not at random assumption. In the left panel, the missing propensity estimated from model (\ref{mmnar}) was plotted against the missing propensity estimated from the following model assuming missing at random.
\begin{equation}
\textup{logit} \{P(R=1)\} = d_0 +d_1 U +d_2 Z.\label{mmar}
\end{equation}
Substantial differences of estimated propensities under the two assumptions were observed.  The largest difference was as high as 0.21, and 24\% of the differences were observed to be greater than 0.1. In the right panel, we plotted the log odds ratio of the estimated propensity from model (\ref{mmar}) against the estimated propensity from model (\ref{mmnar})  versus the gold standard weight. From the figure, we observed a clear decreasing trend of the log odds with the increasing of the true body weight, which implied the difference of the two estimated propensities were related to the true body weight.  For participant with less weight ($< 200$ lb),  model (\ref{mmar}) with missing at random tended to overestimate the  propensity, and for larger weight, model (\ref{mmar}) tended to underestimate the propensity. These results were consistent with the conclusion of missing not at random obtained by both our proposed test and the logistic regression in Table 2.

\begin{table}
	\caption{Model fitting results for logistic regression. The outcome is the indicator for reporting body weight at the day before the month 6 visit, covariates are weight at month 6, baseline weight, and baseline body mass index.}
	\label{table3}
	\centering
	\begin{tabular}{cccc}
		\hline
		& effect size &SE & P-value\\
		\hline
		intercept &0.10&0.15&0.54\\
		$Y^\star$: weight at month 6 &-0.03&0.01&0.02\\
		$U$: baseline weight & 0.03&0.02&0.08\\
		$Z$: body mass index  & 0.06&0.05&0.27\\
		\hline
	\end{tabular}\\
\end{table}

\section{Simulation}
We conducted simulation studies to examine the finite sample behavior of our proposed test. We considered univariate $Y$, $U$ and $Z$ from multivariate normal distribution where $[Y|U,Z] \sim N(1+U+b_zZ,1)$, $[U|Z] \sim N(1-Z,1)$ and $Z\sim N(0,1)$. We generated the missingness indicator variable $R$ for the response variable $Y$ from a Bernoulli distribution with $pr(R=1|Y,U,Z)\sim \Phi(c_0+c_1f(Y)+c_2U)$, where $\Phi(\cdot)$ is the cumulative density for standard normal distribution. By specifying different functional forms for $f(\cdot)$, we allowed different types of dependence of missingness on the value of $Y$. The parameter $b_z$ can be considered as a quantification of how strong the instrumental variable is related to $Y$. In the simulation, we let $b_z = 1$ or $0.5$ to investigate the impact of association between $Z$ and $Y$ on the power of the proposed test. Furthermore, the magnitude of $c_1$ specifies the strength of dependence of missingness on $Y$. The value $c_1=0$ corresponds to the null hypothesis setting where the missingness is at random. We varied the value of $c_1$ to evaluate the size and power for the proposed test. We also chose different values for $c_2$ in order to evaluate the influence of the strength of association between missingness and the variable $U$ on the size and power of the test. An appropriate value of $c_0$ was chosen for each $(c_1, c_2)$ to get an overall missing percentage of $20\%$. Details of the parameter settings are proposed in Table S1 of the Supplementary Material.

Table \ref{table1} showed the estimated levels of Type I error and power from $1000$ replications of the test with sample size of 1000. When the null hypothesis of missing at random is true, i.e. $c_1=0$, most of the rejection rates of the test were within $95\%$ confidence intervals for the nominal levels of $5\%$, i.e. $3.6$--$6.4\%$. A plot of the quantiles of the test statistics against those of the chi-squared distribution indicated that the chi-squared distribution works well at levels other than $5\%$; see Figure S1 in the Supplementary Material. In Table 4, we observed that when sample size increased to $2000$, the test still controlled Type I error at nominal level and obtained higher statistical power. 

The power of the test of missing at random increased with $c_1$, which quantifies the dependence between missingness and response variable. The functional form $f(\cdot)$ has a sizable impact on the power of the proposed test. On the other hand, the magnitude of the parameter $c_2$ has relatively small impact on the power. At the nominal level of $5\%$, the proposed test had about $90\%$ power when $c_1$ was $0.3$ for $f(y)=y$, when $c_1$ was $0.4$ for $f(y)=0.4y^2$, and when $c_1$ was $0.4$ for $f(y)=2.5I(y>1)$. The magnitude of the parameter $b_z$ has a strong impact on the power of the proposed test. When $b_z$ reduced from $1$ to $0.5$, the power also dropped nearly a half for all scenarios.

\begin{table}
	\caption{Empirical rejection rates (\%) of the test for missing at random with sample size $n=1000$ over $500$ replications. The parameter $c_1$ specifies the strength of dependence of missingness on the response variable $Y$, with $c_1=0$ corresponding to the null hypothesis of missing at random. The parameter $c_2$ specifies the strength of association between missingness and covariates $U$. A parameter $c_0$ in each case was chosen to maintain an overall missing percentage of $20\%$. }  
	\centering
	\label{table1}	\scalebox{0.95}{
		\begin{tabular}{ccccccccccccc}
			\hline
			\hline
			\multicolumn{13}{c}{Weak Association ($b_z = 0.5$)}\\
			\hline
			\hline
			&&\multicolumn{11}{c}{$c_1$}\\
			$f(y)$&$c_2$&0&0.05&0.1&0.15&0.2&0.25&0.3&0.35&0.4&0.45&0.5\\
			\hline
			$y$&0 & 5.2 & 7.6 & 12.4 & 17.0 & 27.8 & 37.0 & 55.3 & 60.8 & 72.3 & 84.2 & 88.2 \\ 
			&0.25 & 7.0 & 7.2 & 11.4 & 16.0 & 23.0 & 36.4 & 49.8 & 52.6 & 70.7 & 76.4 & 84.4 \\ 
			&0.50 & 5.8 & 7.6 & 10.4 & 14.0 & 24.4 & 29.0 & 38.6 & 57.8 & 66.8 & 78.2 & 85.6 \\ 
			&0.75 & 4.4 & 6.4 & 9.4 & 11.6 & 20.6 & 28.6 & 39.4 & 49.0 & 54.6 & 68.6 & 82.4 \\ 
			\hline
			$0.4y^2$&0 & 5.0 & 7.8 & 11.4 & 19.0 & 27.8 & 35.8 & 54.8 & 63.9 & 65.2 & 69.6 & 72.6 \\ 
			&0.25 & 4.6 & 6.0 & 9.6 & 13.6 & 19.8 & 29.0 & 27.9 & 35.3 & 45.6 & 45.6 & 55.9 \\ 
			&0.50 & 6.4 & 6.0 & 8.6 & 10.8 & 16.2 & 19.0 & 23.6 & 33.4 & 30.5 & 41.6 & 39.9 \\ 
			&0.75 & 5.4 & 6.4 & 5.2 & 8.4 & 11.4 & 19.2 & 15.6 & 22.6 & 33.4 & 31.0 & 32.2 \\ 
			\hline
			2.5$I(y>1)$&0 & 5.2 & 7.6 & 8.6 & 9.6 & 16.4 & 21.4 & 22.8 & 33.0 & 45.8 & 56.8 & 72.0 \\ 
			&0.25 & 6.0 & 9.2 & 7.8 & 11.2 & 11.4 & 16.4 & 35.0 & 37.0 & 52.5 & 60.4 & 65.4 \\ 
			&0.50 & 6.4 & 6.6 & 6.4 & 11.6 & 16.2 & 23.6 & 30.8 & 40.2 & 46.2 & 56.2 & 68.2 \\ 
			&0.75 & 5.6 & 5.4 & 8.0 & 10.0 & 13.6 & 18.8 & 36.6 & 37.2 & 49.4 & 53.4 & 62.8 \\ 
			\hline
			\hline
			\multicolumn{13}{c}{Strong Association ($b_z = 1$)}\\
			\hline
			\hline
			&&\multicolumn{11}{c}{$c_1$}\\
			$f(y)$&$c_2$&0&0.05&0.1&0.15&0.2&0.25&0.3&0.35&0.4&0.45&0.5\\
			\hline
			$y$&0 & 5.8 & 11.0 & 20.0 & 40.0 & 70.6 & 86.4 & 97.4 & 99.0 & 100.0 & 100.0 & 100.0 \\ 
			&0.25 & 6.6 & 10.2 & 21.6 & 36.8 & 63.4 & 79.4 & 96.6 & 99.2 & 99.8 & 99.8 & 100.0  \\ 
			&0.50 & 7.0 & 7.8 & 18.8 & 37.6 & 57.4 & 74.6 & 89.0 & 96.4 & 99.0 & 99.8 & 100.0 \\ 
			&0.75 & 5.0 & 9.4 & 17.2 & 29.0 & 56.6 & 66.8 & 88.2 & 91.0 & 97.6 & 99.8 & 100.0\\
			\hline
			$0.4y^2$&0 & 7.0 & 13.0 & 34.0 & 60.6 & 82.6 & 95.4 & 96.6 & 98.4 & 99.2 & 99.2 & 100.0  \\ 
			&0.25 & 7.0 & 9.2 & 24.0 & 44.2 & 57.4 & 82.6 & 89.0 & 95.2 & 95.8 & 98.2 & 99.0\\ 
			&0.50 & 4.6 & 6.6 & 17.0 & 35.2 & 50.4 & 56.8 & 77.6 & 87.0 & 90.8 & 95.2 & 97.2 \\ 
			&0.75 & 6.8 & 8.4 & 16.4 & 20.2 & 36.2 & 45.6 & 62.8 & 78.0 & 86.8 & 86.8 & 91.6 \\
			\hline
			2.5$I(y>1)$&  0 & 6.0 & 8.6 & 14.8 & 21.2 & 35.2 & 60.2 & 74.4 & 86.0 & 94.2 & 97.6 & 99.2 \\ 
			&0.25 & 8.0 & 10.4 & 16.0 & 28.8 & 39.6 & 54.8 & 70.8 & 94.6 & 95.6 & 97.8 & 99.8\\ 
			&0.50 & 5.2 & 7.6 & 13.6 & 23.6 & 41.0 & 50.8 & 74.0 & 85.8 & 92.8 & 97.0 & 99.6  \\ 
			&0.75 & 7.0 & 7.2 & 11.6 & 20.6 & 31.6 & 41.8 & 66.8 & 83.6 & 92.0 & 97.2 & 99.2 \\
			\hline
	\end{tabular}}\\
\end{table}

In summary, the proposed test controlled the Type I error well and was reasonably powerful in detecting missing not at random in the settings we considered.

\begin{table}
	\caption{Empirical rejection rates (\%) of the test for missing at random with sample size $n=2000$ over $500$ replications. The parameter $c_1$ specifies the strength of dependence of missingness on the response variable $Y$, with $c_1=0$ corresponding to the null hypothesis of missing at random. The parameter $c_2$ specifies the strength of association between missingness and covariates $U$. A parameter $c_0$ in each case was chosen to maintain an overall missing percentage of $20\%$.  } 
	\label{table2}
	\centering
	\scalebox{0.95}{
		\begin{tabular}{ccccccccccccc}
			\hline
			\hline
			\multicolumn{13}{c}{Weak Association ($b_z = 0.5$)}\\
			\hline
			\hline
			&&\multicolumn{11}{c}{$c_1$}\\
			$f(y)$&$c_2$&0&0.05&0.1&0.15&0.2&0.25&0.3&0.35&0.4&0.45&0.5\\
			\hline
			$y$&0& 7.4 & 8.0 & 14.8 & 28.9 & 44.2 & 60.2 & 76.6 & 90.0 & 97.2 & 99.0 & 99.6 \\ 
			&0.25 & 4.8 & 7.2 & 14.6 & 30.2 & 47.0 & 57.2 & 73.6 & 88.4 & 93.0 & 94.4 & 97.2 \\ 
			&0.5 & 4.2 & 7.2 & 15.0 & 21.0 & 36.0 & 55.0 & 73.8 & 81.4 & 90.4 & 96.4 & 98.0 \\ 
			&0.75 & 2.6 & 7.4 & 10.2 & 20.4 & 39.2 & 46.8 & 64.4 & 83.0 & 91.6 & 95.4 & 98.0 \\ 
			\hline
			$0.4y^2$ & 0 & 5.8 & 10.4 & 24.6 & 36.0 & 56.2 & 73.6 & 79.8 & 83.8 & 85.2 & 94.4 & 96.2 \\ 
			&0.25 & 3.8 & 9.0 & 19.0 & 22.2 & 33.0 & 51.4 & 64.4 & 72.4 & 77.6 & 83.9 & 89.2 \\ 
			&0.50 & 6.2 & 6.8 & 10.2 & 18.4 & 36.8 & 38.4 & 42.0 & 53.0 & 67.0 & 62.6 & 74.0 \\ 
			&0.75 & 5.6 & 3.0 & 8.0 & 15.2 & 22.2 & 31.3 & 37.0 & 47.2 & 55.0 & 58.9 & 60.8 \\ 
			\hline
			2.5$I(y>1)$&0 & 4.4 & 8.6 & 9.6 & 17.2 & 23.0 & 37.4 & 59.2 & 60.4 & 78.1 & 88.0 & 88.4 \\ 
			&0.25 & 6.0 & 5.8 & 12.0 & 16.2 & 28.9 & 34.6 & 50.0 & 66.6 & 77.2 & 85.8 & 92.6 \\ 
			&0.50 & 2.8 & 5.2 & 6.2 & 13.4 & 26.3 & 40.2 & 46.4 & 61.0 & 75.4 & 83.8 & 93.4 \\ 
			&0.75 & 5.2 & 6.8 & 9.2 & 16.0 & 26.4 & 35.8 & 51.4 & 64.0 & 71.3 & 88.4 & 95.8 \\
			\hline
			\hline
			\multicolumn{13}{c}{Strong Association ($b_z = 1$)}\\
			\hline
			\hline
			&&\multicolumn{11}{c}{$c_1$}\\
			$f(y)$&$c_2$&0&0.05&0.1&0.15&0.2&0.25&0.3&0.35&0.4&0.45&0.5\\
			\hline
			$y$&0 & 5.8 & 16.0 & 42.4 & 77.8 & 95.0 & 99.4 & 100.0 & 100.0 & 100.0 & 100.0 & 100.0 \\ 
			&0.25 & 6.4 & 14.0 & 39.2 & 71.4 & 97.2 & 99.4 & 99.8 & 100.0 & 100.0 & 100.0 & 100.0  \\ 
			&0.50 & 6.1 & 13.4 & 33.6 & 64.2 & 91.2 & 97.0 & 100.0 & 100.0 & 100.0 & 100.0 & 100.0\\ 
			&0.75 &4.8& 10.0 & 31.2 & 57.6 & 81.6 & 97.2 & 99.4 & 100.0 & 100.0 & 100.0 & 100.0 \\ 
			\hline
			$0.4y^2$&0 & 5.8 & 22.2 & 64.8 & 92.4 & 98.4 & 99.8 & 100.0 & 100.0 & 100.0 & 100.0 & 100.0  \\ 
			&0.25 & 6.4 & 15.8 & 49.8 & 79.8 & 96.8 & 98.4 & 99.8 & 100.0 & 100.0 & 100.0 & 100.0  \\ 
			&0.50 & 6.8 & 11.4 & 33.8 & 67.6 & 82.4 & 94.2 & 98.6 & 99.4 & 100.0 & 100.0 & 100.0 \\ 
			&0.75 & 8.2 & 10.0 & 24.2 & 53.6 & 75.8 & 89.6 & 97.2 & 98.8 & 99.4 & 99.6 & 99.8\\ 
			\hline
			2.5$I(y>1)$&  0 & 4.2 & 9.8 & 22.2 & 45.8 & 65.6 & 85.0 & 97.8 & 99.2 & 100.0 & 100.0 & 100.0 \\ 
			&0.25 & 6.5 & 13.2 & 20.8 & 44.2 & 65.4 & 85.6 & 98.4 & 99.2 & 99.8 & 100.0 & 100.0  \\ 
			&0.50 & 5.8 & 12.2 & 24.0 & 43.4 & 63.2 & 87.2 & 96.6 & 99.4 & 100.0 & 100.0 & 100.0  \\ 
			&0.75 & 4.0 & 9.8 & 24.2 & 39.0 & 64.4 & 82.2 & 94.6 & 97.7 & 99.8 & 100.0 & 100.0 \\ 
			\hline
	\end{tabular}}\label{table4}\\
\end{table}

\section{Discussion}
In this paper we proposed a new testing procedure to investigate the missing at random assumption for an outcome $Y$ in generalized linear models. We developed a general test statistic in Theorem 1 based on a discrepancy measure of two estimators under two different assumptions, respectively, i.e., missing at random and missing not at random. We provided a realization of the proposed test statistic by choosing the IPW estimator under the missing at random assumption and the estimator proposed by \cite{ZhaoShao_2015_JASA} under the missing not at random assumption constructed {with} the existence of an instrumental variable. Using a newly developed method by \cite{ichimura2015influence}, we derived the influence functions of the two estimators in a semiparametric setting to avoid parametric specification of the missing propensity and the joint distribution of the covariate variables. The realization of the proposed test was validated and evaluated by a simulation study and we found the proposed test was able to control the type I error and provide reasonable statistical power.  {Using data from a weight loss study where 47\% of the at-home body weights of 191 participants were missing},  we investigated the nature of the missing data mechanism using BMI as an instrumental variable. We found strong evidence of missing not at random. Such a finding is consistent with our analysis using validation data where the body weights of all participants were measured a day after the at-home body weights. Our cases study illustrated the practical utility of the proposed testing procedure. The R code has been properly documented and is available online.

The general testing procedure in Theorem 1 covers a broad class of {tests for investigating the missing data mechanism}  and can be extended beyond the scope of the generalized linear models. For example, we can consider the semiparametric density ratio model by \cite{luo2011proportional} which extends the generalized linear model by leaving the reference distribution unspecified. When the outcome is longitudinal,   models proposed in \cite{luo2014moment} and \cite{chen2015regression} can also be considered. While all being flexible, a challenge is to find two estimators and derive their corresponding influence functions,  where one is only valid under missing at random and the other is valid under both missing at random and missing not at random assumptions.

In this paper, the dispersion parameter $\lambda$ is considered as known in our paper. If $\lambda$ is unknown, we define $\theta = (\beta^{T},\lambda)^{T}$, and all the conclusions in this paper still hold with $\beta$ replaced by $\theta$ with corresponding adjustments in the regularity conditions. Alternatively, we may also adapt a two-stage estimation procedure via pseudolikelihood \citep{gong1981pseudo,liang1996asymptotic,chen2010asymptotic}, where $\lambda$ can be replaced by a consistent estimator.  In addition, from the perspective of estimation rather than hypothesis testing, the quantity $(\widetilde \beta - \widehat \beta)/{\widehat \beta}$ can be used to measure the relative bias attributable to the assumption of missing at random when the missingness is truly not at random. This measure can be a helpful complement to a p-value from the proposed testing procedure. Finally, the proposed procedure can be extended to the scenarios with missing covariate variables. Some of the extensions are currently under investigation and will be reported in the future.

\setstretch{1.24}
\bibliographystyle{chicago}
\bibliography{bib-jasa}

\end{document}